\documentclass[nofootinbib,aps,10pt,superscriptaddress,twocolumn]{revtex4-1}

\usepackage[dvips]{graphicx}

\usepackage{siunitx}
\usepackage{array,amsmath,amsthm,feynmf,mathtools}
\usepackage{mathtools}
\usepackage{epsfig}
\usepackage{graphicx}
\usepackage{color}
\usepackage{slashed}
\usepackage{amssymb}

\newcommand{\packageX}{\emph{Package}-{\tt\bfseries X}}

\newcommand{\comm}[1]{{\fontsize{10}{10}\selectfont$\mathtt{#1}$}}

\makeatletter
\let\saved@underbrace\underbrace
\renewcommand*\underbrace[1]{\@ifnextchar_{\ub@with{#1}}{\ub@without{#1}}}
\def\ub@with#1_#2{\mathpalette\underbrace@i{{#1}{_{#2}}}}
\newcommand*\ub@without[1]{\mathpalette\underbrace@i{{#1}{}}}
\newcommand*\underbrace@i[2]{\underbrace@ii#1#2}
\newcommand*\underbrace@ii[3]{\saved@underbrace{#1#2}#3}
\makeatother

\begin{document}

\title{\emph{Package}-X 2.0: A \emph{Mathematica} package for the analytic calculation of one-loop integrals}
\author{Hiren H. Patel}
\email{hiren.patel@mpi-hd.mpg.de}
\affiliation{Particle and Astro-Particle Physics Division \\
Max-Planck Institut fuer Kernphysik {\rm{(MPIK)}} \\
Saupfercheckweg 1, 69117 Heidelberg, Germany}
\begin{abstract}
%
This article summarizes new features and enhancements of the first major update of {\packageX}.  {\packageX} 2.0 can now generate analytic expressions for arbitrarily high rank dimensionally regulated tensor integrals with up to \emph{four} distinct propagators, each with arbitrary integer weight, near an arbitrary even number of spacetime dimensions, giving UV divergent, IR divergent, and finite parts at (almost) any real-valued kinematic point.  Additionally, it can generate multivariable Taylor series expansions of these integrals around any non-singular kinematic point to arbitrary order.  All special functions and abbreviations output by {\packageX} 2.0 supports \emph{Mathematica}'s arbitrary precision evaluation capabilities to deal with issues of numerical stability.  Finally, tensor algebraic routines of {\packageX} have been polished and extended to support open fermion chains both on and off shell.  The documentation (equivalent to over 100 printed pages) is accessed through \emph{Mathematica}'s \emph{Wolfram Documentation Center} and contains information on all {\packageX} symbols, with over 300 basic usage examples, 3 project-scale tutorials, and instructions on linking to \textsc{FeynCalc} and \textsc{LoopTools}.\\

\noindent\textbf{Program summary}\\[2mm]
\emph{Program title:} Package-X\\[2mm]
\emph{Program obtainable from:} CPC Program Library, Queen's University, Belfast, N. Ireland, \emph{or} {\tt http://packagex.hepforge.org}\\[2mm]
\emph{Licensing provisions:} Standard CPC license, {\tt http://cpc.cs.qub.ac.uk/licence/licence.html}\\[2mm]
\emph{Programming language:} Mathematica (Wolfram Language)\\[2mm]
\emph{Operating systems:} Windows, Mac OS X, Linux (or any system supporting Mathematica 8.0 or higher)\\[2mm]
\emph{Journal reference of previous version:} H. H. Patel, Comput. Phys. Commun \textbf{197}, 276 (2015)\\[2mm]
\emph{Does the new version supersede the previous version?:} Yes\\[2mm]
\emph{Summary of revisions:} Extension to four point one-loop integrals with higher powers of denominator factors, separate extraction of UV and IR divergent parts, testing for power IR divergences, construction of Taylor series expansions of one-loop integrals, numerical evaluation with arbitrary precision arithmetic, manipulation of fermion chains, improved tensor algebraic routines, and much expanded documentation.\\[2mm]
\emph{RAM required for execution:} 10 MB, depending on size of computation\\[2mm]
\emph{Vectorised/parallelized?:} No\\[2mm]
\emph{Nature of problem:} Analytic calculation of one-loop integrals in relativistic quantum field theory.\\[2mm]
\emph{Solution method:} Passarino-Veltman reduction formula, Denner-Dittmaier reduction formulae, and additional algorithms described in the manuscript.\\[2mm]
\emph{Restrictions:} One-loop integrals are limited to those involving no more than four denominator factors.\\[2mm]
\emph{Running Time:} 5ms to 10s for integrals typically occurring in practical computations; longer for higher rank tensor integrals.
\newpage
\end{abstract}

\maketitle
\section{Introduction}
{\packageX} is a \emph{Mathematica} package with the principal purpose of generating \emph{analytic} results for dimensionally regulated one-loop rank-$P$ tensor integrals of the form
\begin{multline}\label{eq:tensorintegral}
T_N^{\mu_1\ldots\mu_P}=\Big(\frac{ie^{-\gamma_\text{E} \epsilon}}{(4\pi)^{d/2}}\Big)^{\!-1}\mu^{2\epsilon}\!\!\int\!\! \frac{d^d k}{(2\pi)^d} \Big\{k^{\mu_1} \cdots k^{\mu_P}\\
\times [(k\!+\!p_0)^2\!-\!m_0^2\!+\!i\varepsilon]^{-\nu_0}\,[(k\!+\!p_1)^2\!-\!m_1^2\!+\!i\varepsilon]^{-\nu_1}\\
 \cdots [(k\!+\!p_{N-1})^2\!-\!m_{N-1}^2\!+\!i\varepsilon]^{-\nu_{N-1}}\Big\}\,,
\end{multline}
with up to $N=4$ denominator factors, each with integer weights $\nu_i$, for arbitrary configurations of external momenta $p_i$ and real-valued internal masses $m_i$.

While many packages are publicly available to numerically evaluate one loop integrals \cite{vanOldenborgh:1990yc,Hahn:1998yk,Carrazza:2016gav,vanHameren:2010cp,Cullen:2011kv,Denner:2016kdg, Arbuzov:2016wfy, *Andonov:2004hi} with the aim of automatizing the calculation of cross sections with full kinematic dependence, none exist to the author's knowledge that provide complete analytic expressions.  {\packageX} serves to fill this gap with the aim of obtaining results for quantum field theory calculations where compact analytic expressions exist (\emph{e.g.} pole masses, electroweak oblique parameters, particle moments, decay rates, cross sections at threshold, counterterms, Wilson coefficients, \emph{etc}).  The application files along with an introductory tutorial are hosted at the Hepforge \cite{Buckley:2006nm} project page \texttt{http://packagex.hepforge.org}.

The original release of {\packageX} in 2015 suffered from several limitations, which are listed in Section IX of the accompanying publication \cite{Patel:2015tea}.  In addition to addressing all these limitations, {\packageX} 2.0 features many new features and enhancements described below.  Familiarity with the use of the package is assumed.  Furthermore, since the underlying algorithms and program structure is already detailed at length in the original publication, only a brief description of updates to {\packageX} is provided in the sections to follow.  The interested reader is encouraged to consult the original {\packageX} publication and the references in the text below for details.

The most prominent features of version 2.0 that are described below are the ability to compute one-loop integrals with four distinct denominator factors (Sections \ref{sec:LoopIntegrate}, \ref{sec:LoopRefine} and \ref{sec:ScalarD0}), construct Taylor series expansions of one-loop integrals (Section \ref{sec:LoopRefineSeries}), and compute integrals with open fermion lines (Section \ref{sec:FermionLine}).  Finally, a listing of changes and new functions/symbols are collected in the Appendices.

\section{Updates to \fontsize{10}{8}\texttt{L}\fontsize{8}{8}\texttt{oop}\fontsize{10}{10}\texttt{I}\fontsize{8}{8}\texttt{ntegrate}}\label{sec:LoopIntegrate}
The evaluation of a one-loop integral is initiated with \comm{LoopIntegrate} and performs its covariant tensor decomposition in terms of Passarino-Veltman coefficient functions.
As of version 2.0, \comm{LoopIntegrate} supports integrands with up to \emph{four} distinct propagator factors, each with arbitrary integer powers $\nu_1,\ldots,\nu_{N-1}$.  The tensor covariant decomposition follows the basic property of Lorentz covariance of dimensionally regulated loop integrals \cite{Passarino:1978jh}, and generates an expression in terms of coefficient functions \comm{PVA}, \comm{PVB}, \comm{PVC}, \comm{PVD} as described in section III of \cite{Patel:2015tea}.

In the original version of {\packageX}, the syntax of \comm{LoopIntegrate} was such that the user had to put the integral in a form such that $p_0 = 0$ in (\ref{eq:tensorintegral}).  This meant that the user either had to strategically route the momenta through the one-loop diagram so that at least one denominator factor had no external momenta flowing thorough it, or had to shift the integration variable before using \comm{LoopIntegrate} to evaluate the integral.  As of version 2.0, \comm{LoopIntegrate} accepts a new syntax accommodating nonzero $p_0$ which means that any momentum routing is possible.

For integrals involving denominator factors with linearly dependent momenta, it is standard practice to perform a partial fraction expansion of the denominator factors before making a covariant decomposition as it leads to a sum of integrals with fewer number of denominator factors.  As of version 2.0, \comm{LoopIntegrate} has an option \comm{Apart} that toggles whether to expand linearly dependent denominator factors into partial fractions.  The algorithms are based on \cite{Feng:2012iq}, but are specifically tailored to efficiently process one-loop integrals.

Additionally, if the numerator of the loop integral contains dot products involving the loop momentum ($k^2$ or $k.p_i$), it is profitable to write them as differences of propagator factors so that they may be cancelled against the denominator.  This procedure simultanesouly lowers the rank of tensor integrals and reduces the number of denominator factors.  As of version 2.0, \comm{LoopIntegrate} has an option \comm{Cancel} that controls whether to expand and cancel these factors before making a covariant decomposition.  This, along with expanding in partial fractions, generally leads to substantially increased performance and quality of the output of \comm{LoopRefine}.

Finally, in version 2.0, \comm{LoopIntegrate} can now perform the covariant decomposition for integrands involving \comm{DiracMatrix}, \comm{FermionLine}, and \comm{FermionLineProduct} objects, enabling it to process integrands with open fermion lines.  This capability is described in Section \ref{sec:FermionLine} below.

\section{Updates to \fontsize{10}{8}\texttt{L}\fontsize{8}{8}\texttt{oop}\fontsize{10}{10}\texttt{R}\fontsize{8}{8}\texttt{efine}}\label{sec:LoopRefine}
After the covariant decomposition of a loop integral is carried out with \comm{LoopIntegrate} and kinematic conditions are supplied, the final step is to apply \comm{LoopRefine}, which replaces the coefficient functions with explicit analytic expressions.

In version 2.0, \comm{LoopRefine} can convert four-point coefficient functions \comm{PVD} to elementary functions.  In the most general case, the standard Passarino-Veltman reduction \cite{Passarino:1978jh} is used.  To handle cases with vanishing Grammian determinant, the algorithms described in \cite{Denner:2005nn} are used, and are direct generalizations of \emph{Cases 1, 3, 5} and \emph{6} appearing in section IV C of \cite{Patel:2015tea} to four propagator factors.  These algorithms are not applicable if the determinant of the modified Cayley matrix vanishes.  For further (but still incomplete) coverage, two new reduction formulae, valid when certain elements of the adjugate Grammian matrix vanish, were derived by explicitly integrating over Feynman parameters.  When appropriate, {\packageX} applies these formulae.  However, further research is needed to provide complete kinematic coverage.


In the original release of {\packageX}, only integrals with unit weight were supported ($\nu_0=\ldots=\nu_{N-1}=1$).  As of version 2.0, \comm{LoopRefine} can convert integrals of arbitrary integer weight to analytic expressions.  Weighted loop integrals can arise when Feynman diagrams with massless gauge bosons in general covariant gauge are considered, or when the small momentum approximation is applied at the level of the integrand.  They are also indispensable for the construction of series expansions as described in Section \ref{sec:LoopRefineSeries} below.  Conventionally, integration by parts methods \cite{Chetyrkin:1981qh} together with recurrence relations in $d$ dimensions \cite{Davydychev:1991va, Tarasov:1996br} are used to reduce these functions.  However, {\packageX} uses a slightly different method derived as follows: by comparing Feynman parameter integral representations, the weighted coefficient functions can be related to linear combinations of unweighted coefficient functions with fewer number of $00$ index pairs.  Then one of the existing reduction formulae can be used to convert them to analytic expressions.  However, weighted scalar functions in this scheme are related to coefficient functions with formally negative number of $00$ pairs.  But since their Feynman parameter integral representations imply that they are analytic functions of ``number of $00$ pairs'', a reduction based on Cayley determinants presented in Section 5.3 of \cite{Denner:2005nn} is used to raise the number of $00$ index pairs, thereby making it possible to convert them to known basis functions.

Similar to weighted coefficient functions, coefficient functions arising from the decomposition of integrals near an even number of spacetime dimensions other than 4 can also be related to those defined near 4 with a different number of $00$ index pairs by equating Feynman parameter integral representations.  Therefore, it is possible to convert these functions to analytic expressions by using the same strategy.

In addition to the aforementioned major changes, \comm{LoopRefine} has also received a number of minor features.  Among them is the ability for \comm{LoopRefine} to test for the presence of power infrared divergences in loop integrals in some cases.  These integrals have the feature that their Feynman parameter integrals do not converge for genuinely small dimensional regulator parameter $\epsilon$.  Their presence is detected in {\packageX} by retaining, in all basis functions that feature the power divergence, the $+i\varepsilon$ term as a finite quantity which acts like a mass regulator.  Then, upon converting its entire input to analytic expressions, \comm{LoopRefine} checks whether the $+i\varepsilon\rightarrow0$ limit in the total is well-behaved.  This accommodates the possibility that the power infrared divergence cancels among various terms in the input expression.  Furthermore, \comm{LoopRefine} takes an additional option \comm{Analytic}, which when set to \comm{True}, generates a result with the parameter $\epsilon$ analytically continued to large and negative values, rending power infrared divergences formally convergent.

Finally \comm{LoopRefine} can restrict its computation to just the infrared or ultraviolet divergent parts by appropriately setting the option \comm{Part} to \comm{UVDivergent} or \comm{IRDivergent}, respectively.  In addition to providing a quick way to obtain the $1/\epsilon$ pole parts of loop integrals, this new feature also enables one to identify their origin as either ultraviolet or infrared.  The calculation of the ultraviolet divergent part is fast owing to the existence of an iterative formula applicable to an arbitrary coefficient function \cite{Sulyok:2006xp}.  The calculation of the IR divergent part is sped up by setting to zero all coefficient functions with at least one pair of $00$ indices since they are known to be infrared finite \cite{Denner:2005nn}.

\section{The scalar $D_0$ function: analytic expressions and numerical implementation}\label{sec:ScalarD0}
The reduction algorithm for converting coefficient $D$-functions, \comm{PVD}, in the case of non-vanishing Grammian determinant ends with the UV-finite scalar function $D_0$.  To complete the computation of the one-loop integral, \comm{LoopRefine} replaces this scalar function.

In order for \comm{LoopRefine} to faithfully display the $1/\epsilon$ poles in the final output, all IR divergent cases are substituted.  Expressions with massive internal lines are drawn from \cite{Ellis:2007qk} and \cite{Denner:2010tr}, and those with massless internal lines are adapted from \cite{Duplancic:2000sk}.  Care has been taken to ensure that all analytic expressions are consistent with the $+i\varepsilon$ prescription so that their numerical evaluation yields the correct sign for their imaginary parts.

The situation for the IR finite four-point functions is unlike the case of the three- and lower- point scalar functions in that no reasonably compact analytic expressions are known unless the Grammian determinant vanishes.  For these cases, \comm{LoopRefine} simply outputs \comm{ScalarD0}, with the function itself implemented numerically.  The principle behind the numerical implementation of the scalar four-point function follows that of the three-point function outlined in Section V of \cite{Patel:2015tea}.  In the region of positive modified Cayley determinant (which covers the physical region), the imaginary part is obtained by applying Cutkosky's rule, and requires the evaluation of a single logarithm for each channel above normal threshold.  The real part is based on a representation \cite{Nierste:1991a, Denner:1991qq} in terms of sixteen numerically evaluated dilogarithm functions \cite{Osacar:1995aa}.  Outside this region, the separation of the real and imaginary parts is not used since the application of Cutkosky's rule is more complicated.  For rapid numerical evaluation, the code for \comm{ScalarD0} is compiled to the \emph{Wolfram Virtual Machine}.

Additionally, as of version 2.0, every special numerical function available in {\packageX} (including \comm{ScalarD0}) is written in terms of native \emph{Wolfram Kernel} functions so that the user may take advantage of \emph{Mathematica}'s arbitrary precision evaluation capabilities.  This new feature of {\packageX} allows one to obtain numerically stable results for kinematic configurations that would otherwise lead to severe loss of precision.

If an analytic expression is desired, \comm{ExpandD0} can be applied to \comm{ScalarD0}, which replaces it with the corresponding analytic formula from {\packageX}'s library.  The simpler expressions are adapted from \cite{Davydychev:1993ut} and \cite{Duplancic:2002dh}, and the most general and complicated cases are derived from \cite{Denner:2010tr}.  In order to minimize the size of the output, the analytic expressions are given compactly as \comm{RootSum}s over four Denner-Beenakker continued dilogarithms $\mathcal{L}i_2(x,y)$ defined originally in \cite{Beenakker:1988jr} and implemented in {\packageX} as \comm{ContinuedDiLog}.

\section{\fontsize{10}{8}\texttt{L}\fontsize{8}{8}\texttt{oop}\fontsize{10}{10}\texttt{R}\fontsize{8}{8}\texttt{efine}\fontsize{10}{8}\texttt{S}\fontsize{8}{8}\texttt{eries}: Taylor series expansions of loop integrals}\label{sec:LoopRefineSeries}
Calculations within the standard model and its extensions usually involves integrals with widely separated scales.  In these cases, an exact analytic representation of these integrals are too verbose to be of any value, and their numeric evaluation usually suffers from loss of precision due to large numerical cancellations.  As a result, one usually desires an approximate expression obtained from the first few terms of its series expansion.

{\packageX} 2.0 provides a way to obtain Taylor series expansions of loop integrals with a new routine \comm{LoopRefineSeries}, which is to be used in place of \comm{LoopRefine}.  Internally, the algorithm for constructing a series is as follows: 
\begin{enumerate}
\item Differentiate the Passarino-Veltman coefficient functions as needed to generate the necessary terms in the Taylor expansion.
\item Express the differentiated coefficient functions in terms of undifferentiated ones with fewer number of $00$ index pairs.
\item Apply the appropriate reduction formula on the functions to convert the expression to analytic form.
\end{enumerate}
In step 2, the conventional practice is to relate the differentiated coefficient functions to those defined in a lower number of spacetime dimensions \cite{Tarasov:1996br}.  But as  described in Section \ref{sec:LoopRefine}, it is more convenient to use the standard coefficient functions with fewer pairs of $00$ indices, and apply the reduction formula based on Cayley determinants in step 3  for negative pairs.  This method of constructing series expansions easily generalizes to multiple series expansions, and to arbitrarily high order, constrained only by memory and computation time.

A major limitation of this method is that it is unable to construct expansions around Landau singularities because the necessary derivatives at those points usually do not exist.  Therefore, it is not possible to construct small mass, threshold, or other \emph{asymptotic} expansions with \comm{LoopRefineSeries} alone.  While one can still circumvent this problem in limited circumstances by applying \emph{Mathematica}'s \comm{Series} function on the output of \comm{LoopRefine}, a fully automated routine to construct such expansions is absent.

\section{Support for open fermion lines}\label{sec:FermionLine}
Although the primary task of {\packageX} is to assist in the analytic calculation of one-loop integrals of the form given in (\ref{eq:tensorintegral}), the kinds of integrals one encounters when calculating in physical theories are those involving fermions.  The original release of {\packageX} included the function \comm{Spur} to calculate traces of Dirac matrices, and satisfactorily handled Feynman integrals involving a closed fermion loop.  However, no direct support for evaluating integrals with \emph{open} fermion lines was provided.  In order to evaluate these integrals, the integrals had to be projected onto form factors one at a time.  For calculations involving several Feynman integrals, projecting onto separate form factors for each one is laborious.  In order to alleviate this problem, version 2.0 now provides direct support for open fermion lines.

The object \comm{DiracMatrix} represents a product of Dirac matrices, arising from an off shell fermion line in a Feynman diagram.  Additionally, two new algebraic objects are introduced in version 2.0: \comm{FermionLine} represents a product of Dirac matrices sandwiched on either side by on shell spinors, and \comm{FermionLineProduct} represents the direct product of \comm{FermionLine} objects.  These new objects can be used for integrals involving one or more open fermion lines in a given diagram.  They are algebraically manipulated by \comm{LoopIntegrate} and  \comm{FermionLineExpand} to bring them to canonical form by following a series of steps:
\begin{enumerate}
\item Expand the product of Dirac matrices by distributing multiplication over addition.
\item Relate products of gamma matrices with repeated Lorentz indices (such as $\gamma_\mu\gamma^\nu\gamma^\rho\gamma^\mu$) to products with fewer gamma matrices.
\item Apply the Dirac algebra bringing $\slashed{p}$ to either end of the fermion line in order to apply the Dirac equation.
\item Apply the Sirlin identities \cite{Sirlin:1981pi} to doubly-contracted direct products of Dirac matrices (\comm{FermionLineProduct} only).
\item Resolve remaining products of Dirac matrices into SVTAP basis \cite{Chisholm:1971tu}.
\item Apply the Gordon identities \cite{Gordon:1928a} to express
    the vector and axial-vector convective transition currents in terms of Dirac/Pauli and Anapole/EDM transition currents.
\end{enumerate}
The on shell relations in step 3 and 5 are not used for manipulating \comm{DiracMatrix} objects since they are not applicable.  Additionally, when manipulating  \comm{FermionLineProduct} objects, the entire algorithm is repeated until the identities can be no longer applied.  Note that since $\gamma_5$ is implemented naively in dimensional regularization, any algebraic manipulations involving $\gamma_5$ is valid in exactly 4 dimensions only.  Likewise, the identities used in steps 4 and 5 are valid in exactly 4 dimensions.  Therefore, care should be taken to ensure that the integrals are either manifestly finite, or have been properly regulated to ensure correctness of the finite part of the result.

\section{Expansion of the Documentation}
Included with the package software is a set of documentation files accessible from within \emph{Mathematica}.  The documentation contains information for all front end functions and available options in {\packageX}.  In version 2.0, the documentation is expanded to include more details and usage examples for greater clarity.  Additionally, three project-scale tutorials are included illustrating how {\packageX} may be used in research, briefly summarized below:
\begin{itemize}
\item \emph{Ward Identities and $\gamma_5$ in Dimensional Regularization} -- In {\packageX}, $\gamma_5$ is naively defined to anti-commute with all other gamma matrices, and may lead to incorrect results for logarithmically divergent integrals.   This tutorial illustrates, with the $Z^*\gamma\gamma$ Green function as an example, how chiral Ward identities can be enforced using Alder's method \cite{Jegerlehner:2000dz, Adler:1970}.
\item \emph{Extracting Form Factors from the $\mu\rightarrow e\gamma$ amplitude} -- This tutorial explains the use of \comm{Projector} in detail, while also explaining how {\packageX} can be used to verify the decoupling theorem, gauge invariance, and Ward identities. 
\item \emph{Scattering of Light by Light} -- This tutorial explains how to compute the light-by-light scattering matrix element at leading order, and illustrates the capabilities and limitations of covariant methods used in {\packageX}.
\end{itemize}
The documentation also contains instructions on linking {\packageX} to two other publicly available packages: {\sc FeynCalc} \cite{Mertig:1990an, Shtabovenko:2016sxi} (through \textsc{FeynHelpers} \cite{Shtabovenko:2016whf}) and to {\sc LoopTools} \cite{Hahn:1998yk}.

\section{Summary and avenues for further development}
In this paper, new features of {\packageX} in the most recent major release is described.  Version 2.0 significantly expands the scope of the software by supporting four point integrals, Taylor expansions, and open fermion lines.   However, there are still many limitations of {\packageX}  which continue to guide its development summarized below:

\begin{enumerate}
\item Currently, {\packageX} is unable to automatically construct asymptotic expansions of loop integrals around Landau singularities.  Since such expansions are very relevant in quantum field theory calculations (\emph{e.g.} for mass regularization, threshold expansions, eikonal expansions), an update including this feature is desirable.

\item Additionally, the reduction algorithms for the four-point functions do not completely cover the case when the Cayley determinant vanishes.  These cases become physically relevant when the Grammian determinant also vanishes, since they can correspond to IR divergent scattering amplitudes at physical threshold.  It would be ideal to provide complete kinematic coverage for these singular cases as well.

\item As {\packageX} is applied to larger problems, manual input of all integrals required for a specific calculation becomes increasingly tedious and prone to input errors.  It would be desirable to have a way, \emph{e.g.} by linking to the \textsc{FeynArts} package \cite{Hahn:2000kx}, to generate all the needed integrals in {\packageX} automatically.

\item For large problems, it may not be possible to simplify the integrals to obtain compact analytic expressions.  Although not its primary purpose, it would be convenient to be able to directly evaluate the coefficient functions numerically, without needing their analytic expressions.  While linking to \textsc{LoopTools} \cite{Hahn:1998yk} is described in the documentation, linking to more modern packages like \textsc{Collier} \cite{Denner:2016kdg} is desirable.

\item {\packageX} is limited to the calculation of Lorentz covariant Feynman integrals.  However, non-covariant integrals are also frequently encountered, for example, when one is working in Coulomb or Axial gauges, or within effective field theories like NRQED, HQET and SCET.  The ability to automatically obtain analytic results for these cases would be convenient.

\end{enumerate}

\begin{acknowledgements}
I would like to thank the numerous testers who participated during the alpha and beta phases of development.  In particular, I thank Michael Duerr, Federica Giacchino, Matthew Kirk, Fatima Machado, Pedro Malta, Kevin Max, Johannes Welter, and Yibo Yang for checking results, exposing bugs, and reporting typos in the program. In addition I thank Sebastian Ohmer and Moritz Platscher for proof-reading the tutorial, and Jiang-Hao Yu for proof-reading this manuscript.  Special thanks goes to Josh Ellis, Julian Heeck, and Tanja Geib for providing truly extensive feedback that helped made {\packageX} better.  I am especially grateful to Vladyslav Shtabovenko, the lead developer of \textsc{FeynCalc}, for stimulating discussions while {\packageX} was being developed, and for providing helpful information about \textsc{FeynHelpers}.

I also thank Ansgar Denner and Stefan Dittmaier for clarifying discussions regarding their work, and I am indebted to Ulrich Nierste for providing me his Master's thesis and explaining his method of analytic continuation of the scalar four-point function in detail.  Finally, I thank Jakub Kuczmarski, a member of the \emph{Mathematica} StackExchange community, for answering countless questions about \emph{Wolfram Workbench} and his add-on \mbox{\textsc{WWBCommon}}, which helped to make the built-in documentation possible.
\end{acknowledgements}

\appendix
\section{Convention and package structure changes between versions 1.0 and 2.0}
\begin{itemize}
\item \comm{pvA}, \comm{pvB}, \comm{pvC}, \comm{pvC0} and \comm{pvC0IR6} are now named \comm{PVA}, \comm{PVB}, \comm{PVC}, \comm{ScalarC0} and \comm{ScalarC0IR6}, respectively, consistent with \emph{Mathematica}'s naming convention of capitalizing the first letter of pre-defined symbols.  Dimensional regularization 't Hooft scale \comm{\mu R} is renamed to \comm{\mu} ({\texttt{[\textbackslash Micro]}}).

\item \comm{LoopIntegrate} now normalizes its integration measure so that $e^{\gamma_E \epsilon}$  is factored out instead of $r_\gamma = \frac{\Gamma^2(1-\epsilon)\Gamma(1+\epsilon)}{\Gamma(1-2\epsilon)}$.  This was done to prepare {\packageX} for computing two loop integrals in the future.  With respect to version 1.0, this change in version 2.0 \emph{only} modifies the output of \comm{LoopRefine} exhibiting $1/\epsilon^2$ poles, arising from \emph{overlapping} soft and collinear IR divergences, by an amount proportional to $-\pi^2/12$ to the finite part.

\item The order of arguments of Passarino-Veltman $C$ functions have been changed to match that of other popular packages such as {\sc FeynCalc,  LoopTools, Collier}, \emph{etc}, including other authors in the literature. The relations are
\begin{multline*}
\qquad\mathtt{PVC[r,n1,n2,s1,s12,s2,m0,m1,m2]}=\\
\mathtt{pvC[r,n1,n2,s1,s2,s12,m2,m1,m0]}
\end{multline*}\\[-8mm]and\\[-8mm]
\begin{multline*}
\qquad\mathtt{ScalarC0[s1,s12,s2,m0,m1,m2]}=\\
\mathtt{pvC0[s1,s2,s12,m2,m1,m0]}\,.
\end{multline*}

\item Contexts \comm{X`IndexAlg`}, \comm{X`Spur`}, \comm{X`OneLoop`} are deprecated; all package symbols now belong to a common context \comm{X`}.

\item Auxiliary Passarino-Veltman function \comm{pvb} is now obsolete, and is covered by by higher weight \comm{PVB} functions.

\item Off shell fermion self energy form factor $C(p^2)$ has been redefined in \comm{Projector} without a factor of $i$ in front.

\end{itemize}

\section{New functions/symbols introduced in version 2.0}

\begin{itemize}
\item\comm{LoopRefineSeries}\\[2mm]
generates a (multiple) Taylor series expansion of one loop tensor integrals.

\item\comm{PVD}\\[2mm]
represents the Passarino-Veltman tensor coefficient function $D_{\underbrace{0\ldots0}_{2r}\underbrace{1\ldots1}_{n_1}\underbrace{2\ldots2}_{n_2}\underbrace{3\ldots3}_{n_3}}$, and substituted with an analytic expression by \comm{LoopRefine}

\item\comm{ScalarD0}, \comm{ScalarD0IR12}, \comm{ScalarD0IR13}, \comm{ScalarD0IR16}\\[2mm]
gives the finite parts of the scalar four point function $D_0$ for real external invariants and real positive masses, as classified in \cite{Ellis:2007qk}.

\item\comm{C0Expand} and \comm{D0Expand}\\[2mm]
expands out scalar functions \comm{ScalarC0}, \comm{ScalarD0} and related functions in terms of analytic expressions.

\item\comm{Kibble\phi}\\[2mm]
gives the Kibble kinematic polynomial $\phi(s_1,\ldots,s_6)$ \cite{Kibble:1960zz} (the Grammian determinant associated with four particle processes).

\item\comm{ContinuedDiLog}\\[2mm]
gives the Beenakker-Denner continued dilogarithm function $\mathcal{L}i_2(x,y)$ \cite{Beenakker:1988jr}.

\item\comm{MandelstamRelations}\\[2mm]
gives a list of replacement rules expressing Lorentz scalar products in terms of Mandelstam invariants and masses.

\item\comm{LScalarQ}\\[2mm]
allows declaring symbols as being Lorentz scalars.

\item \comm{FermionLine} and \comm{FermionLineProduct}\\[2mm]
represents products of Dirac matrices sandwiched between on shell $u$ and $v$ spinors, and their direct products.

\item \comm{FermionLineExpand}\\[2mm]
expands out and performs the Dirac algebra on \comm{FermionLine}, and \comm{FermionLineProduct} objects to put them in canonical SVTAP form.
\end{itemize}

\bibliography{packageX.bib}

\end{document}